\documentclass[12pt]{article}

\usepackage[a4paper,left=30mm,right=20mm,top=20mm,bottom=20mm]{geometry}
\usepackage{graphics}
\usepackage{amsmath}
\usepackage{epsfig}
\usepackage{cite}
\usepackage{xcolor}
\usepackage{ulem}
\usepackage[english]{babel}

\graphicspath{{figures/}}

\title{Charm fragmentation and associated $J/\psi + Z/W^\pm$ production at the LHC}

\author{S.P.~Baranov$^1$, A.V.~Lipatov$^{2,3}$, A.A.~Prokhorov$^{2,3,4}$ }

\begin{document}

\maketitle

\begin{center}

{\it $^{1}$P.N.~Lebedev Institute of Physics, Moscow 119991, Russia}\\
{\it $^{2}$Skobeltsyn Institute of Nuclear Physics, Lomonosov Moscow State University, 119991, Moscow, Russia}\\
{\it $^{3}$Joint Institute for Nuclear Research, 141980, Dubna, Moscow region, Russia}\\
{\it $^{4}$Faculty of Physics, Lomonosov Moscow State University, 119991 Moscow, Russia}\\

\end{center}

\vspace{0.5cm}

\begin{center}

{\bf Abstract }

\end{center}

\indent

We consider the production of electroweak $Z$ or $W^\pm$ bosons associated with $J/\psi$ 
mesons at the LHC conditions. Our attention is focused on new partonic subprocesses
which yet have never been considered in the literature, namely, the charmed or strange
quark excitation subprocesses followed by the charmed quark fragmentation $c\to J/\psi+c$.
Additionally we take into account the effects of multiple 
quark and gluon radiation in the initial and final states.
We find that the contributions from the new mechanisms are important and
significantly reduce the gap between the theoretical and experimental 
results on the $J/\psi + Z$ and $J/\psi + W^\pm$ production cross sections.

\vspace{1.0cm}

\noindent{\it Keywords:} heavy quarkonia, electroweak bosons,
QCD evolution, small-$x$, TMD parton densities in a proton

\newpage

\section{Motivation} \indent

Hadronic production of $J/\psi$ mesons in association with electroweak gauge bosons
($Z$ or $W^\pm$) are interesting processes\cite{1,2}. They involve both strong and weak 
interactions and may serve as a complex test of perturbative QCD, electroweak theory and
parton evolution dynamics.
Moreover, they provide a unique laboratory to investigate the 
charmonia production mechanisms predicted by the non-relativistic QCD (NRQCD) 
factorization\cite{3,4}. The latter is a rigorous 
framework for the description of heavy quarkonia production and/or decays and
implies a separation of perturbatively 
calculated short distance cross sections for
the production of a heavy quark pair in an intermediate 
Fock state ${}^{2S+1}L_J^{(a)}$ with spin $S$, 
orbital angular momentum $L$, total angular momentum $J$ and 
color representation $a$ from its subsequent non-perturbative transition 
into a physical quarkonium via soft gluon radiation.
However, the charmonia production at high energies is not fully understood at the moment 
despite many efforts made in last decades.
In fact, NRQCD has a long-standing challenge in the $J/\psi$ and $\psi^\prime$ 
polarization and provides inadequate description of the $\eta_c$ production 
data\footnote{One possible solution has been proposed recently, see\cite{5} and references 
therein.} (see, for example,\cite{6,7,8} for more information).
The associated production of $J/\psi$ mesons and 
gauge bosons at the LHC is a suitable process for further studying NRQCD.

A complete next-to-leading order (NLO) predictions for prompt
$J/\psi + Z/W^\pm$ production within the NRQCD are available\cite{10,11,12}.
It was shown that the differential cross sections at the 
leading order (LO) are significantly enhanced by the NLO corrections.
There are only color octet (CO) contributions 
to the $J/\psi + W^\pm$ production at both LO and NLO level\cite{10,11},
while color singlet (CS) terms contribute at higher orders\cite{10}.
Taken together with the corresponding predictions from the 
double parton scattering (DPS) mechanism, the 
NLO NRQCD expectations for associated $J/\psi + Z$ production cross sections
are lower than the data\cite{1} by a factor of $2$ to $5$ (depending on the
$J/\psi$ transverse momentum). The difference between the theoretical predictions 
and the measured $J/\psi + W^\pm$ cross sections is typically even larger\cite{2}.
Of course, these discrepancies need an explanation.
So, further theoretical studies are still an urgent and important task.

In the present paper, we draw attention to a new contributions
to prompt $J/\psi + Z/W^\pm$ production cross section, namely, the 
flavor excitation subprocesses {charm quark excitation for $Z$ bosons (or strange quark 
excitation for $W^\pm$ bosons) followed by subsequent charm fragmentation, $c\to J/\psi + c$.
These contributions have been overlooked in the literature and yet have never been considered.
Of course, this flavor excitation is of formally non-leading order in $\alpha_s$ in the
pQCD expansion, but it could play a role because of its different momentum dependence 
(in comparison with what was previously considered \cite{10,11,12}).
The main goal of our study is to clarify this point.
In addition, we calculate contributions coming from multiple gluon radiation 
in the initial state and final states through CO gluon fragmentation
(via $g\to c\bar c [^3S_1^{(8)}] \to J/\psi$ channel). 
This mechanism was found to be important \cite{13} for double $J/\psi$ production 
in the kinematic region covered by the CMS and ATLAS measurements.
We see certain interest in examining this kind of contributions for other processes,
such as vector boson production in association with $J/\psi$ mesons. 

The outline of the paper is the following. 
In Section~2 we describe the basic steps of our calculations. 
In Section~3 we present the numerical results and discussion.
Our conclusions are summarised in Section~4.

\section{The model} \indent

As it was already mentioned above, the new subprocesses can be described
as the sea (charm or strangeness) excitation followed by the $c$-quark fragmentation:
\begin{gather}
  g + c \to Z + c, \quad g + s \to W^- + c, \quad c \to J/\psi + c,
  \label{fragmentation}
\end{gather}
\noindent
and similar subprocesses for $\bar c$ and $\bar s$ antiquarks. There is no double 
counting with the previous estimations\cite{10,11,12} based on the quark-antiquark annihilation and gluon-gluon fusion subprocesses
\begin{gather}
  q + \bar q \to Z/W^\pm + J/\psi, \quad g + g \to Z/W^\pm + J/\psi,
  \label{NRQCD}
\end{gather}
\noindent
since the subprocesses~(\ref{fragmentation}) have rather different final state containing
$c$-quark (as it is clearly seen in Fig.~\ref{fig:NRQCD}).
Despite non-leading in $\alpha_s$, the subprocesses (\ref{fragmentation}) have 
some advantage in kinematics. The large mass $m$ of the emitted boson in~(\ref{NRQCD})
suppresses two quark propagators and leads to the dependence 
$\sigma \sim 1/m^8$, see Fig.~\ref{fig:NRQCD} (left diagram). 
In contrast, the behavior of quark excitation 
subprocesses~(\ref{fragmentation}) 
is $\sigma \sim 1/m^4$. This compensates the sparseness of the charmed or strange
sea and the presence of extra coupling $\alpha_s$.

\begin{figure}
\begin{center}
{\includegraphics[width=0.5\textwidth]{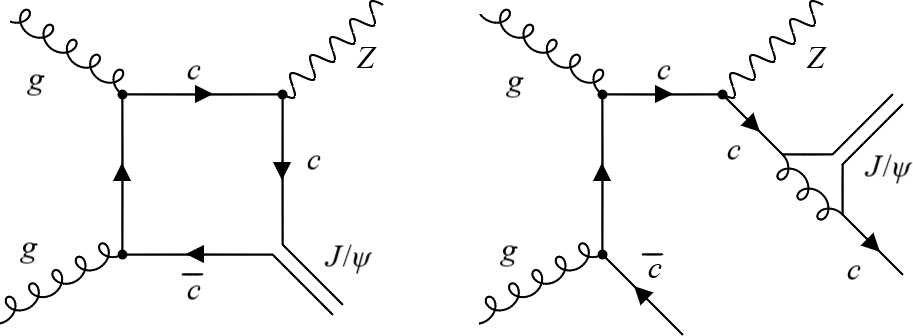}}\hfill
\caption{Example of Feynman diagram taken into account in the NRQCD calculations \cite{10,11,12} (left panel) and diagram of charm excitation followed by the $c$-quark fragmentation to $J/\psi$ (right panel).}
\label{fig:NRQCD}
 \end{center}
\end{figure}

In the calculations shown below, we exploit the idea that the sea quarks all appear from 
a perturbative chain as a result of the QCD evolution of gluon densities in a proton. 
Then, appending an explicit gluon splitting vertex to~(\ref{fragmentation}), 
we come to the gluon-gluon fusion subprocesses, see Fig.~\ref{fig:diagrams} ({\it a}) and ({\it b}):
\begin{gather}
  g + g \to Z + c + \bar c, \quad g + g \to W^- + c + \bar s, \quad c \to J/\psi + c.
  \label{ggV}
\end{gather}
\noindent
The $W^+ + \bar c + s$ production subprocesses can be obtained by charge conjugation.
The difference between~(\ref{fragmentation}) and~(\ref{ggV}) is that the 
sea quark density in~(\ref{fragmentation}) may contain a non-perturbative component.

Yet another class of new contributions is represented by the initial and final state gluon
radiation that accompanies the production of electroweak bosons, where the radiated gluons
then convert into $J/\psi$ mesons via CO channel $g\to c\bar c[^3S_1^{(8)}]\to J/\psi$
(see also\cite{13}).
Here we take into account the following subprocesses:
\begin{gather}
  g + g \to Z + q + \bar q, \quad g + g \to W^\pm + q + \bar{q^\prime}, 
   \label{light_ggV} 
\end{gather}
\noindent
where $q$ denotes any light quark flavor. The initial and final state parton emission
is simulated using the standard parton showering algorithms (see discussion below).
To evaluate the subprocesses (\ref{ggV}) and~(\ref{light_ggV})
we employ the $k_T$-factorization approach\cite{14,15}, which can be considered 
as a convenient alternative to explicit fixed-order perturbative QCD calculations 
at high energies. We see certain technical advantages in the fact that, even with 
the leading-order (LO) amplitudes for hard partonic scattering,
one can include a large piece of higher-order (NLO + NNLO + ...) corrections
(the ones connected with the multiple initial state gluon emission)
taking them into account in the form of transverse momentum dependent (TMD, or unintegrated) 
gluon densities in a proton\footnote{A detailed 
description of this approach can be found, for example, in review\cite{16}.}.
An essential point is making use of the Catani-Ciafaloni-Fiorani-Marchesini (CCFM)\cite{17} 
equation to describe the QCD evolution of gluon distributions. This equation smoothly 
interpolates between the small-$x$ Balitsky-Fadin-Kuraev-Lipatov (BFKL)\cite{18} gluon dynamics 
and high-$x$ Dokshitzer-Gribov-Lipatov-Altarelli-Parisi (DGLAP)\cite{19} dynamics, thus 
providing us with the suitable tool for our phenomenological study.
The gauge-invariant off-shell production amplitudes for subprocesses~(\ref{ggV}) 
and~(\ref{light_ggV}) have been calculated in\cite{20,21} and all of the technical 
details are explained there\footnote{These amplitudes are implemented in the Monte-Carlo 
event generator \textsc{pegasus}\cite{22}.}.
To reconstruct the CCFM evolution ladder, we generate a Les Houches Event file using 
the Monte-Carlo event generator \textsc{pegasus}\cite{22} and then process the file with 
the TMD parton shower routine implemented into the Monte-Carlo event generator 
\textsc{cascade}\cite{23}.
In this way we obtain full information about gluon emissions in the initial state.

\begin{figure}
\begin{center}
{\includegraphics[width=0.7\textwidth]{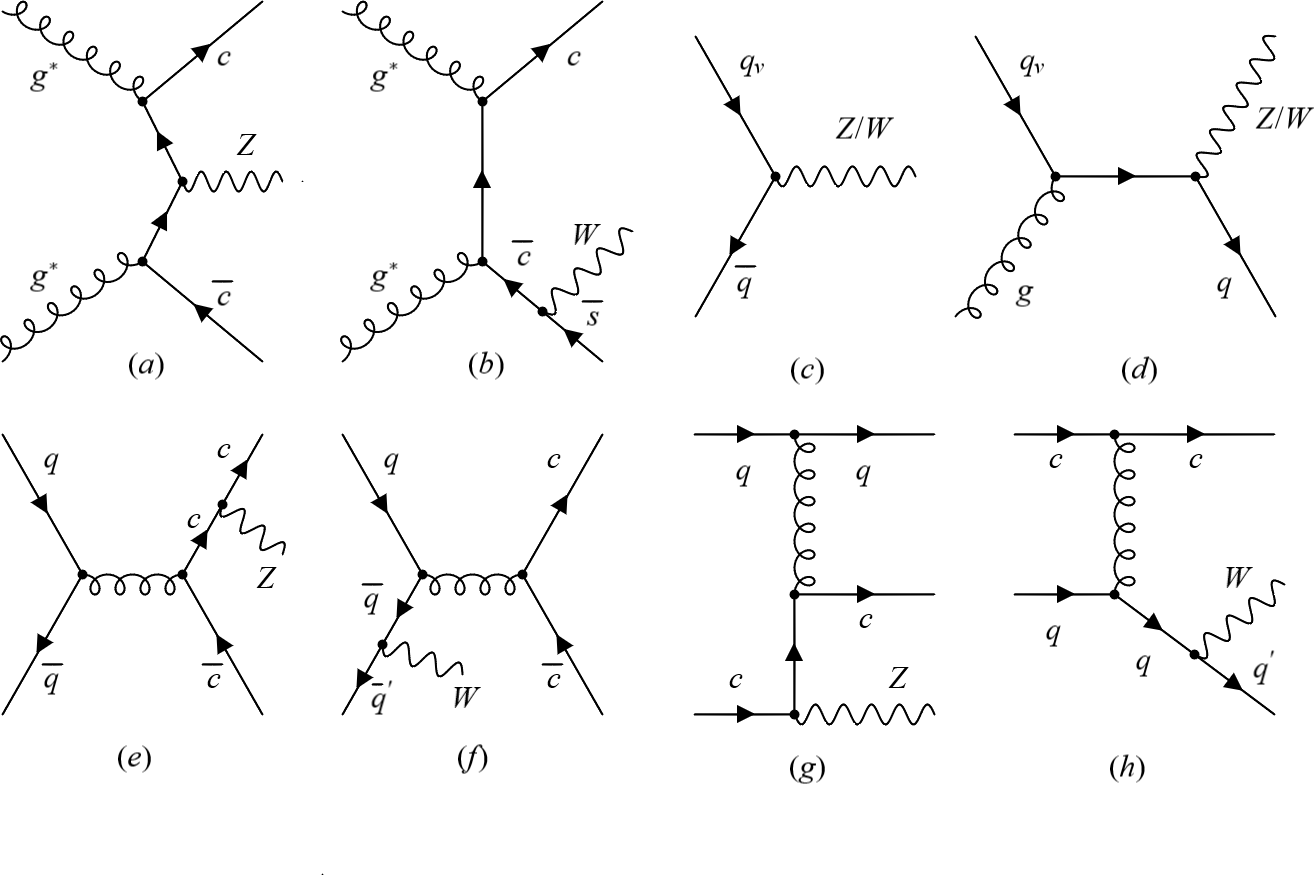}}\hfill
\caption{Examples of Feynman diagrams for off-shell gluon-gluon fusion~(\ref{ggV}) or~(\ref{light_ggV}) 
and quark-induced subprocesses~(\ref{collV}).}
\label{fig:diagrams}
 \end{center}
\end{figure}


For completeness, we also take into account several subprocesses
involving quarks in the initial state shown in Fig.~\ref{fig:diagrams} ({\it c}) --- ({\it h}):
\begin{gather}
q + \bar q \to Z/W^{\pm}, \enskip q + g \to Z/W^{\pm} + q, 
\enskip q + c \to Z + q + c, \enskip q + q^\prime \to W^\pm + q + c,
  \label{collV} 
\end{gather}  
\noindent
where we keep only valence quarks in the first two cases to avoid double counting
with off-shell gluon-gluon fusion subprocesses (\ref{light_ggV}), see
Fig.~\ref{fig:diagrams} ({\it c}) and ({\it d}).
The last subprocesses, shown in Fig.~\ref{fig:diagrams} ({\it e}) - ({\it h}), are taken into account since 
they provide additional charmed quarks, despite they are obviously suppress in $\alpha_s$.
The quark-induced contributions~(\ref{collV}) can play a role at essentially large transverse 
momenta (or, respectively, at large $x$ which is needed to produce high $p_T$ events)
where the quarks are less suppressed or can even dominate over the gluon density.
Here we find it reasonable to rely upon the conventional (collinear) DGLAP-based 
factorization, which provides better theoretical grounds in the region of large $x$. 
So, our scheme represents a combination of two techniques
with each of them being used at the kinematic conditions where it is best 
suitable\footnote{A similar scanario has been succesfully applied\cite{24,25}
to describe the associated $Z + b$ and $Z + c$ production at the LHC (see also\cite{26}).}.
The evaluation of the relevant hard scattering amplitudes and production 
cross sections is straightforward and needs no explanation. 
To simulate the parton emissions in initial and final states, we employ the parton shower 
routine implemented into the Monte-Carlo event generator \textsc{pythia8}\cite{27}.
In this way, we collect all the radiated gluons and charmed quarks.

The parton-level amplitudes (\ref{ggV}) --- (\ref{collV}) provide the starting point for the 
charmed quark and the gluon fragmentation. The reliability of the fragmentation approach 
can be motivated by the high transverse momenta typical for the recent ATLAS data\cite{1,2} 
taken at $\sqrt{s} = 8$~TeV (see, for example, \cite{cfragm,cfragm+}). 

The fragmentation function $\mathcal{D}^{\mathcal{H}}_{a}(z,\mu^2)$ describes the transition 
of a parton $a$ into a charmonium state $\mathcal{H}$ and can be expressed as a collection
of contributions from the different intermediate states:
\begin{gather}
\mathcal{D}^{\mathcal{H}}_{a}(z,\mu^2) =\sum_{n}d^n_a(z,\mu^2)\langle\mathcal{O}^{\mathcal{H}}[n]\rangle,
\end{gather}
\noindent
where $n$ labels the intermediate CS or CO state of the charmed
quark pair, $\mu^2$ is the fragmentation scale and $\langle\mathcal{O}^{\mathcal{H}}[n]\rangle$
are the corresponding long-distance matrix elements (LDMEs)\cite{3,4}. 
Then, the eventual cross section reads
\begin{gather}
\frac{d\sigma(pp\to \mathcal{H} + Z/W^\pm)}{dp_{T}} = \sum_n \int \frac{d\sigma(pp\to c+Z/W^\pm)}{dp^{(c)}_{T}}\,
\mathcal{D}^{\mathcal{H}}_{c}(z,\mu^2) \delta(z-p/p^{(c)}) dz + \nonumber \\ 
+ \sum_n \int \frac{d\sigma(pp\to Z/W^\pm+g)}{dp^{(g)}_{T}}\,
\mathcal{D}^{\mathcal{H}}_{g}(z,\mu^2) \delta(z-p/p^{(g)}) dz,
\end{gather} 
\noindent
where $p^{(c)}$, $p^{(g)}$ and $p$ are the momenta of the $c$-quark, gluon 
and outgoing charmonium state $\mathcal{H}$, respectively.
Only the $c\to c\bar c [^3S^{(1)}_1] + c$ and $g \to c\bar{c}[ ^3S^{(8)}_1]$ transitions give 
sizeable contributions to the $S$-wave charmonia ($J/\psi$, $\psi^\prime$).
For $P$-wave mesons $\chi_{cJ}$ with $J = 0$, $1$ or $2$ one has to take into account the
leading terms $c \to c\bar{c}[ ^3P^{(1)}_J] + c$ and $g \to c\bar{c}[ ^3S^{(8)}_1]$.
In our numerical calculations we take into account the feeddown contributions from 
$\chi_{c1}$, $\chi_{c2}$ and $\psi^\prime$ decays and neglected the $\chi_{c0}$ decays 
because of low branching fraction of the latter\cite{28}.

The scale dependence of the fragmentation functions is driven by the multiple gluon
radiation. We take the initial conditions in the form (see, for example,\cite{29})
\begin{gather}
 d^{[ ^3S^{(8)}_1]}_g(z,\mu^2_0) = \frac{\alpha_s(\mu^2_0)}{m^3_c}\frac{\pi}{24}\delta(1-z), \\
 d^{[ ^3S^{(1)}_1]}_c(z,\mu^2_0) = \frac{\alpha^2_s(\mu^2_0)}{m^3_c}\frac{16z(1-z)^2}{243(2-z)^6}(5z^4-32z^3+72z^2-32z+16),\\
 d^{[ ^3P^{(1)}_1]}_c(z,\mu^2_0) = \frac{\alpha^2_s(\mu^2_0)}{m^5_c}\frac{64z(1-z)^2}{729(2-z)^8} \times\nonumber \\ \times (7z^6-54z^5+202z^4-408z^3+496z^2-288z+96),\\
 d^{[ ^3P^{(1)}_2]}_c(z,\mu^2_0) = \frac{\alpha^2_s(\mu^2_0)}{m^5_c}\frac{128z(1-z)^2}{3645(2-z)^8} \times\nonumber \\ \times(23z^6-184z^5+541z^4-668z^3+480z^2-192z+48),
\end{gather}
\noindent
where the starting scale is $\mu^2_0 = m^2_{\psi}$. The effects of final state multiple
radiation can be described by the LO DGLAP evolution equation 
\begin{gather}
\frac{d}{d \,\ln\mu^2}\begin{pmatrix}\mathcal{D}_c^{\mathcal H}  \\ \mathcal{D}_g^{\mathcal H}  \end{pmatrix} = \frac{\alpha_s(\mu^2)}{2\pi} \begin{pmatrix}P_{qq} ~ P_{gq}   \\ P_{qg} ~ P_{gg} \end{pmatrix}\otimes \begin{pmatrix}\mathcal{D}_c^{\mathcal H} \\ \mathcal{D}_g^{\mathcal H}  \end{pmatrix},
\label{FFevo} 
\end{gather}
\noindent
where $P_{ab}$ are the standard LO DGLAP splitting functions.
According to the NRQCD approximation, we set the charm mass to $m_c = m_{\mathcal{H}}/2$
and then solve the DGLAP equations~(\ref{FFevo}) numerically.

Concerning other parameters involved into our calculations,
we have used the TMD gluon densities in a proton 
obtained from the numerical solution of CCFM evolution equation, namely, 
JH'2013 set 1 and set 2 gluons\cite{30}. 
The input parameters of JH'2013 set 1 gluon distribution have been fitted to the 
precise HERA measurement of the proton structure function $F_{2}(x,Q^2)$, whereas the
input parameters of JH'2013 set 2 gluon were fitted to the 
both structure functions $F_{2}(x,Q^2)$ and $F_{2}^c(x,Q^2)$.
According to\cite{30}, we use the two-loop formula for the QCD coupling $\alpha_s$ with $n_f = 4$ active 
quark flavors at $\Lambda^{(4)}_{\rm QCD} = 200$~MeV and 
factorization scale $\mu^2_F = \hat s + \mathbf{Q}^2_T$, 
where $\hat s$ is the total energy of partonic subprocess 
and $\mathbf{Q}_T$ is the transverse momentum of initial off-shell gluon pair. 
This choice is dictated by the CCFM evolution algorithm. 
For quark-induced subprocesses we choose the MMHT'2014 (LO)\cite{31} set
and apply one-loop expression for $\alpha_s$ with  $n_f = 5$ quark flavors at $\Lambda^{(5)}_{\rm QCD} = 211$~MeV.
Everywhere we set the renormalization and fragmentation 
scales, $\mu_R$ and $\mu_{fr}$, to be equal to half of the sum of 
transverse masses of produced particles and transverse mass of fragmented parton, respectively.
The CS LDMEs of $J/\psi$ and $\psi^\prime$ mesons are known from their 
decay widths: $\langle\mathcal{O}^{J/\psi}[ ^3S^{(1)}_1]\rangle = 1.16$~GeV$^3$, 
$\langle\mathcal{O}^{\psi^\prime}[ ^3S^{(1)}_1]\rangle = 0.7038$~GeV$^3$ (see, for example,\cite{3,4} and references therein).
The CS wave functions at the origin for $\chi_{cJ}$
mesons and CO LDMEs for charmonia family have been determined\cite{32,33}: 
$\langle\mathcal{O}^{\chi_{c1}}[ ^3P^{(1)}_1]\rangle = 0.2$~GeV$^5$, 
$\langle\mathcal{O}^{\chi_{c2}}[ ^3P^{(1)}_2]\rangle = 0.0496$~GeV$^5$, 
$\langle\mathcal{O}^{J/\psi}[ ^3S^{(8)}_1]\rangle = 0.0012$~GeV$^3$,
$\langle\mathcal{O}^{\psi^\prime}[ ^3S^{(8)}_1]\rangle = 0.0012$~GeV$^3$
and 
$\langle\mathcal{O}^{\chi_{c0}}[ ^3S^{(8)}_1]\rangle = 0.0004$~GeV$^3$.
Masses and branching fractions of all particles involved into the calculations 
were taken according to Particle Data Group\cite{28}.

\section{Numerical results} \indent

In this section, we present the results of our calculations 
and perform a comparison with the recent ATLAS data\cite{1,2}.
The ATLAS Collaboration has measured the differential cross sections 
of associated $Z/W^\pm + J/\psi$ production
as a function of $J/\psi$ transverse momentum in a 
restricted part of the phase space (fiducial volume) at $\sqrt{s} = 8$~TeV. 
In the case of $Z + J/\psi$ production, the $J/\psi$ meson was required to
have transverse momentum $p^{J/\psi}_T > 8.5$~GeV and rapidity
$|y^{J/\psi}| < 2.1$, while
the leading and subleading muons originated from subsequent Z boson decays must have 
pseudorapidities $|\eta^{l}| < 2.5$
and transverse momenta $p^{l}_T > 25$~GeV and $p^{l}_T > 15$~GeV, respectively.
Their invariant mass $m_{ll}$ is required to be $|m_{ll} - m_Z| <$~10 GeV, where $m_Z$
is the $Z$ boson mass.
For $W^\pm + J/\psi$ production, the following cuts are applied:
$p^{J/\psi}_T > 8.5$~GeV, $|y^{J/\psi}| < 2.1$, $p^{l}_T > 25$~GeV and $|\eta^{l}| < 2.4$ for 
muon and $p^{\nu}_T > 20$~GeV for neutrino originated from the $W^\pm$ boson decays.
The transverse mass of $W^\pm$ boson defined as 
\begin{gather}
  m_T(W^{\pm})=\sqrt{2p_T^l p_T^\nu[1-\cos(\phi^{l}-\phi^{\nu})]}
\end{gather} 
\noindent
is required to be $m_T(W^{\pm}) > 40$~GeV, where 
$\phi^{l}$ and $\phi^{\nu}$ are the azimuthal angles of the decay muon and neutrino.
We have implemented the experimental setup used by the ATLAS Collaboration 
in our calculations.

\begin{figure}
\begin{center}
{\includegraphics[width=.5\textwidth]{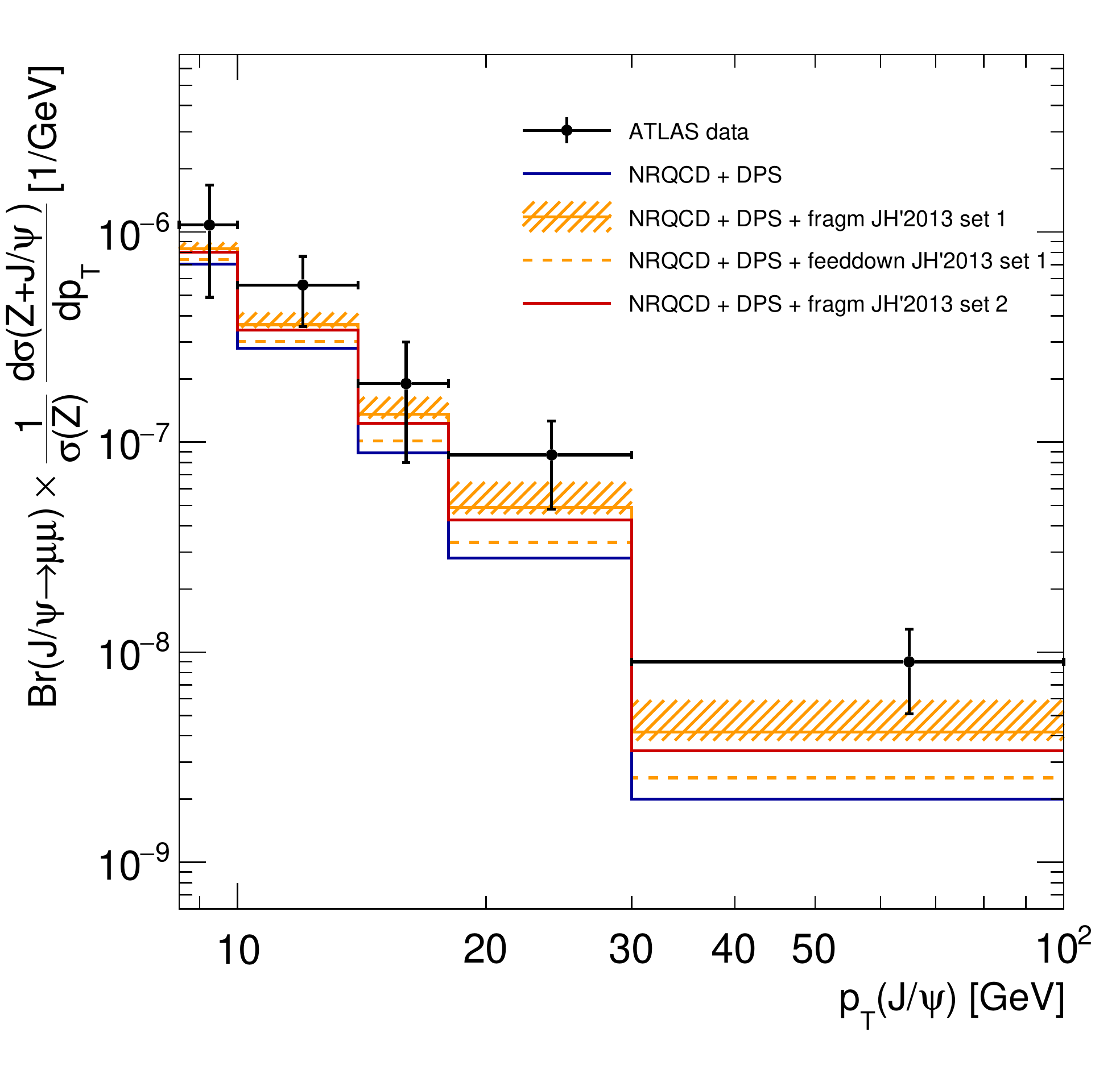}}\hfill
{\includegraphics[width=.5\textwidth]{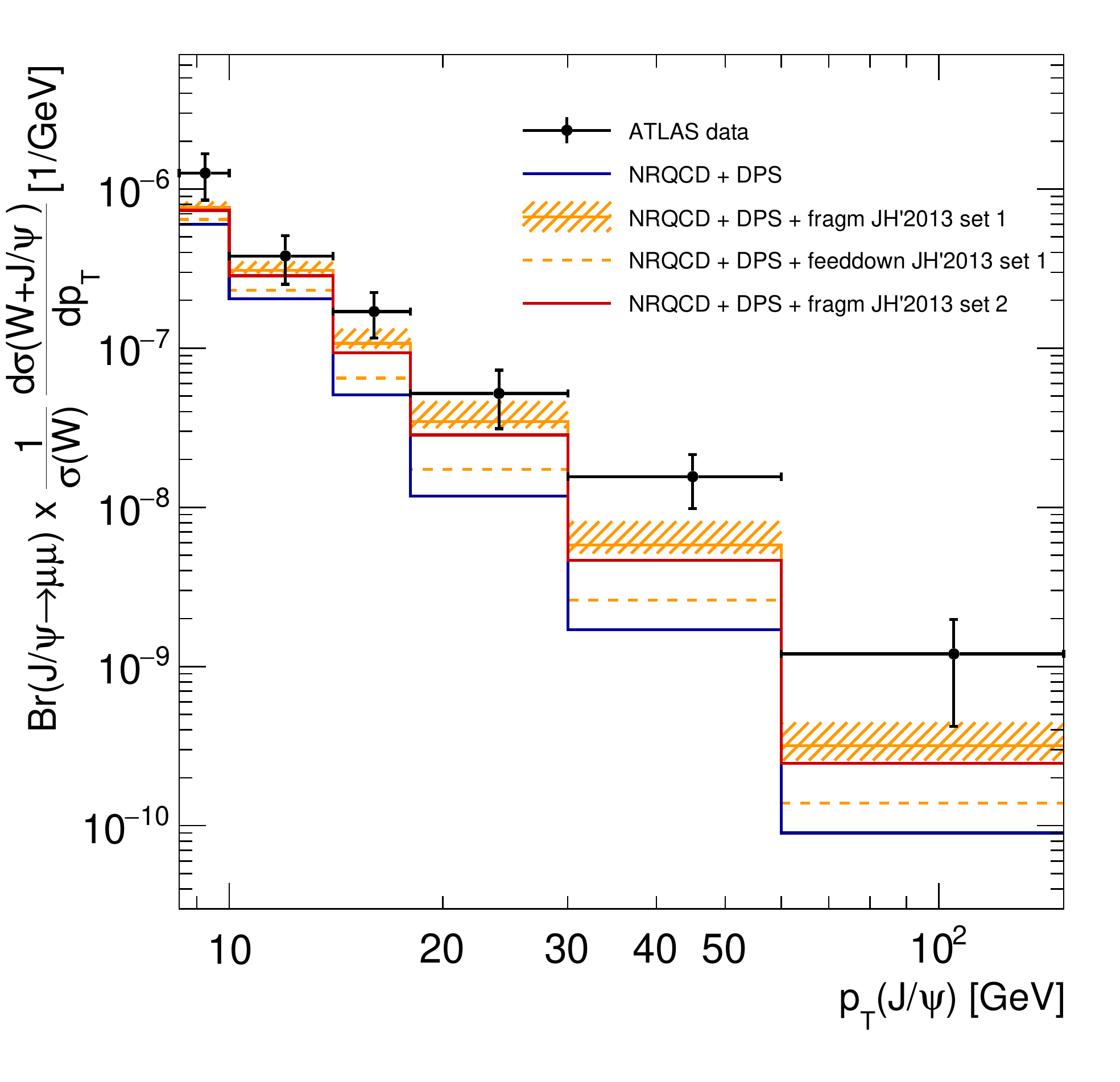}}\hfill
\caption{The differential cross section of  associated $Z + J/\psi$ (left panel) and $W^{\pm} + J/\psi$ (right panel) 
production in $pp$ collisions at $\sqrt{s} = 8$~TeV. The kinematical cuts applied are described in the text.
The NLO NRQCD + DPS predictions are taken from\cite{1,2}. The DPS contribution was estimated 
with $\sigma_{\rm eff} = 15$~mb. Experimental data are from ATLAS\cite{1,2}.
The uncertainty band shown in the figure includes only the uncertainties from the original
newly calculated subprocesses (\ref{ggV}) --- (\ref{collV}) and not from (\ref{NRQCD}). }
\label{fig:ZWJpsi}
 \end{center}
\end{figure}

Our numerical results are shown in Fig.~\ref{fig:ZWJpsi}, where 
we have used JH'2013 set 1 gluon density as the default choice.
We find that the contributions from subprocesses~(\ref{ggV}) --- (\ref{collV})
with their subsequent parton fragmentation into $J/\psi$ mesons
are remarkably important, especially at large transverse momenta.
In fact, at $p^{J/\psi}_T \geq 20 - 30$~GeV it gives approximately the same contribution as 
that coming from the NLO NRQCD estimations summed with the corresponding contribution from 
the DPS production mechanism (we took the latter from the ATLAS papers\cite{1,2}).
The contributions~(\ref{ggV}) --- (\ref{collV}) are large for $W^\pm + J/\psi$ 
production (where only the CO contributions are presented in the NLO NRQCD predictions) 
almost in the whole range of $p^{J/\psi}_T$.
One can see that summing the contributions from subprocesses~(\ref{ggV}) --- (\ref{collV})
and NLO NRQCD predictions (and DPS terms, of course) allows us to significantly reduce 
the discrepancy between the theoretical expectations and experimental data. Moreover, 
the upper edge of our estimated uncertainty band shown in Fig.~\ref{fig:ZWJpsi}
is rather close to the ATLAS data for both processes under consideration.

As usual, to evaluate the uncertainties we have varied the renormalization and fragmentation scales around their default values by a factor of $2$. We only note that we replaced the JH'2013 set 1 gluon density by the JH'2013 set 1$+$
or JH'2013 set 1$-$ ones when calculating the uncertainties connected with
the variation of renormalization scale in~(\ref{ggV}) and~(\ref{light_ggV}).
This was done to preserve the intrinsic consistency of the calculation, that is, to observe
the correspondence between the TMD gluon set and scale used in the CCFM evolution 
(see\cite{30} for more information).
The dominant uncertainties are found to come from the scales of gluon densities, while the ones
coming from fragmentation functions are almost negligible compared to the first.

To investigate the sensitivity of our results to the choice of 
TMD gluon density we repeat the calculations with JH'2013 set 2 distribution.
We find that the obtained predictions are rather close to the 
ones obtained with the default JH'2013 set 1 gluon (or even 
coincide with them within the uncertainties).
The feeddown contributions from $\psi^\prime$ and $\chi_{cJ}$ decays 
also play a significant role. 
Their contribution is about of $30$\% of the estimated direct
contribution in a wide $p^{J/\psi}_T$ range, as one can see in Fig.~\ref{fig:ZWJpsi}.

\begin{figure}
\begin{center}
{\includegraphics[width=.5\textwidth]{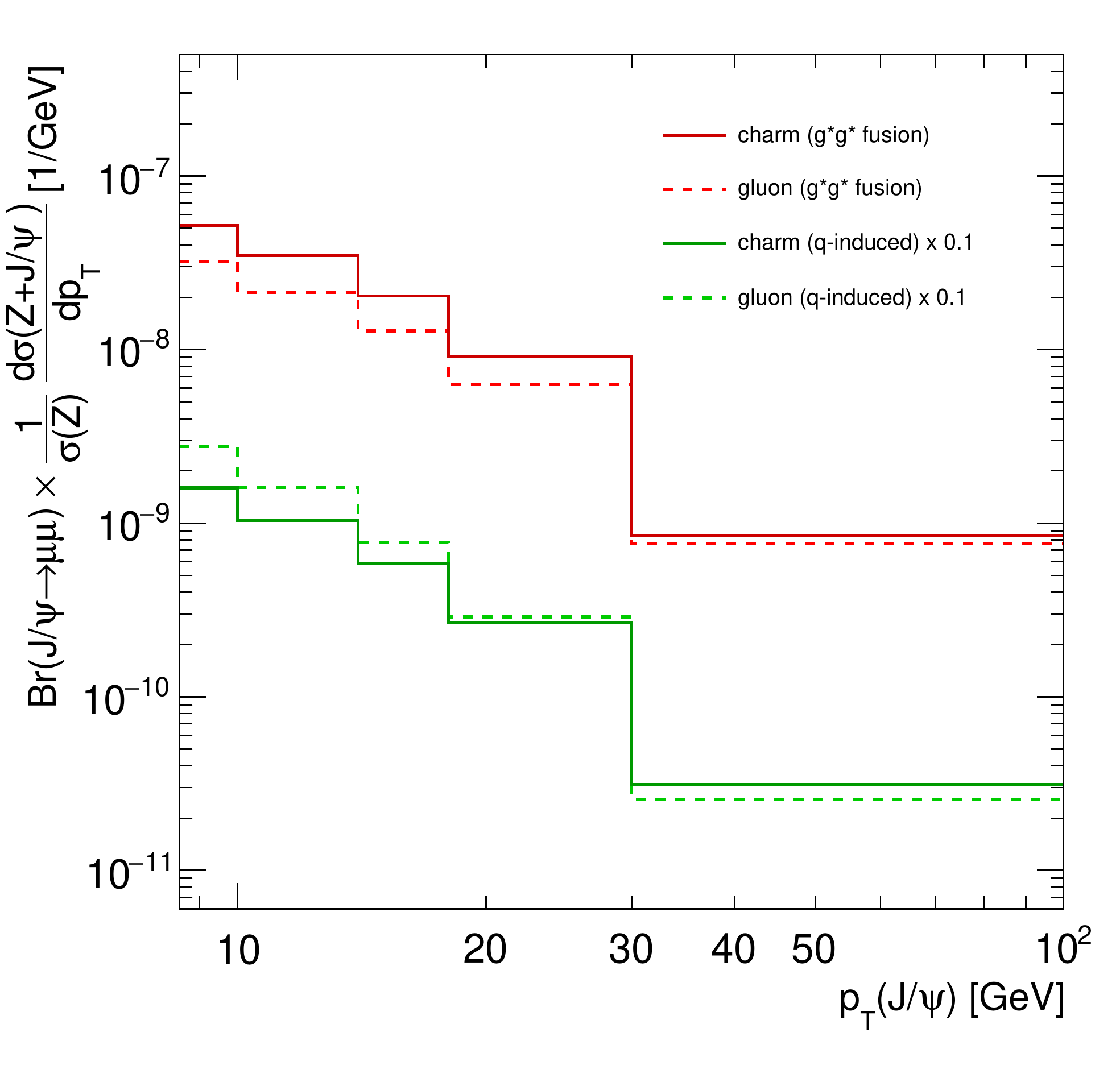}}\hfill
{\includegraphics[width=.5\textwidth]{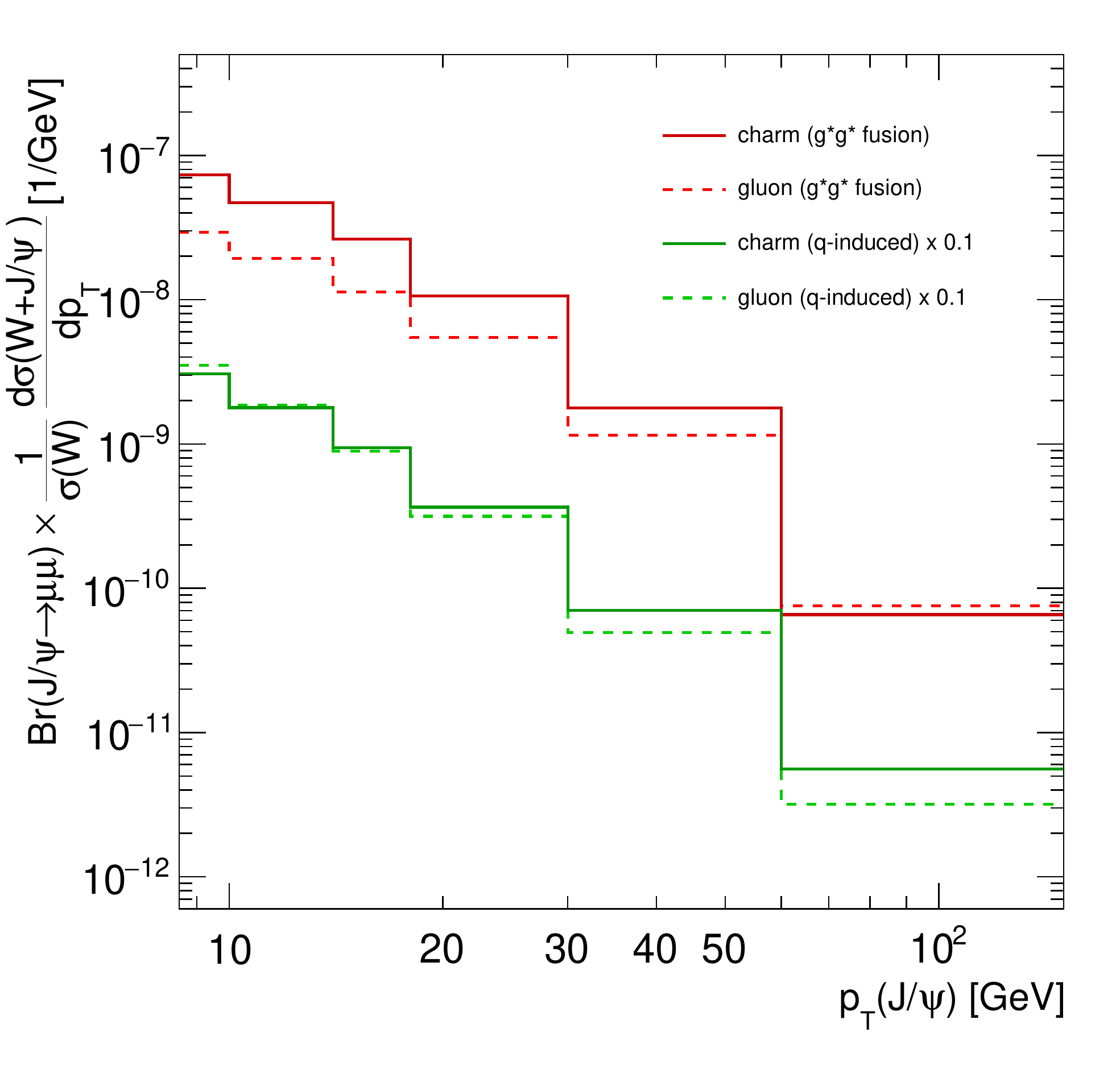}}\hfill
\caption{Contributions from charm and gluon fragmentation to the 
associated $Z + J/\psi$ (left panel) and $W^{\pm} + J/\psi$ (right panel) production in $pp$ collisions at $\sqrt s = 8$~TeV. Contributions come from the off-shell gluon-gluon fusion  correspond to JH'2013 set 1 gluon density.}
\label{fig:fragm_mechanisms}
 \end{center}
\end{figure}

Now let us discuss the role of multiple parton radiation. 
As it was mentioned above, we considered here two qualitatively different 
sources of parton fragmentation into the $J/\psi$ mesons, namely, fragmentation of charmed 
quarks, originated in the hard interaction and 
fragmentation of gluons, originated as a result of initial
QCD evolution of parton densities.
Note that the charmed quarks emitted in the initial state 
according to \textsc{pythia} parton showering algorithm
were attributed to the charm quark fragmentation to keep the charm/gluon separation. 
The effects of parton showers in the final states are taken
into account in the form of DGLAP-evolved fragmentation functions.
The contributions from charmed quarks and gluon fragmentation are shown in Fig.~\ref{fig:fragm_mechanisms} separately. 
In the case of gluon-gluon fusion subprocesses,
the main contribution at the small transverse momenta $p^{J/\psi}_T$ comes 
from fragmentation of charmed quarks while at the large $p^{J/\psi}_T$ 
the gluon fragmentation starts to dominate.
For quark-induced subprocesses~(\ref{collV})
small $p^{J/\psi}_T$ region is driven mainly by the fragmentation of multiple gluon radiation,
whereas the region of high $p^{J/\psi}_T$ is governed by the charm quark fragmentation.
Nevertheless, as one can clearly see, in both cases the fragmentation of multiple gluon 
emission noticeably enhances the charm fragmentation and provides a sensible growth of 
the total and differential cross sections. 

In summary, we can conclude that adding the new production 
mechanisms~(\ref{ggV}) --- (\ref{collV}) 
to the conventional NLO NRCQD predictions
greatly improves the situation for prompt $Z/W^\pm + J/\psi$ production 
and significantly decreases the gap between the theoretical estimations and the data. 
Further on, an appropriate treatment of the initial and final state parton emission
and including the feeddown from the $\chi_c$ and $\psi'$ decays enhances the theoretical 
expectations even more and makes the agreement with the data even better.

\section{Conclusion} \indent

We have considered the production of electroweak $Z$ or $W^{\pm}$ bosons associated with 
$J/\psi$ mesons in $pp$ collisions at the LHC at $\sqrt{s} = 8$ TeV. 
We have investigated the role of new partonic subprocesses which yet have never been 
considered in the literature, namely, the flavor (charm or strangeness) excitation 
subprocesses followed by the charm fragmentation $c\to J/\psi + c$.
In addition, we have taken into account the effects of multiple quark and gluon radiation
in the initial and final states.
We have demonstrated that the considered new contributions are remarkably 
important and significantly reduce the gap between the theoretical 
and experimental results on the $J/\psi + Z/W^\pm$ production cross sections.

\section*{Acknowledgements} \indent

The authors thank M.A. Malyshev for very useful discussions and 
careful reading of the manuscript. We thank also
S.M.~Turchikhin for discussion of \textsc{pythia} parton shower 
routine and its implementation to ATLAS setup. 
We are grateful to DESY Directorate for the support in the framework of
Cooperation Agreement between MSU and DESY on phenomenology of the LHC processes
and TMD parton densities. 
A.A.P. was supported by grant 
of the foundation for the advancement of theoretical physics and
mathematics "Basis" 18-2-6-129-1 and 
RFBR grant 20-32-90105.

\end{document}